\documentclass[conference]{IEEEtran}
\IEEEoverridecommandlockouts
\usepackage{listings}
\usepackage{xcolor}
\usepackage[flushleft]{threeparttable}
\usepackage{enumitem} 
\usepackage{tabularx}
\usepackage{booktabs}
\usepackage[numbers, sort&compress]{natbib}  
\usepackage{hyperref}
\hypersetup{
    colorlinks = true,
    linkcolor = blue,
    citecolor = blue,
    filecolor = magenta,
    urlcolor = magenta,
}

\usepackage{amsmath,amssymb,amsfonts}
\usepackage{algorithmic}
\usepackage{graphicx}
\usepackage{textcomp}
\usepackage{xcolor}
\usepackage[utf8]{inputenc}
\usepackage[acronym]{glossaries}
\usepackage{orcidlink}
\def\BibTeX{{\rm B\kern-.05em{\sc i\kern-.025em b}\kern-.08em
    T\kern-.1667em\lower.7ex\hbox{E}\kern-.125emX}}
\begin{document}

\title{Automotive Speed Estimation: Sensor Types and Error Characteristics from OBD-II to ADAS}

\author{
    \IEEEauthorblockN{
        Hany Ragab\IEEEauthorrefmark{1}\,\orcidlink{0000-0003-3167-9100},~\IEEEmembership{Member,~IEEE}, 
        Sidney Givigi\IEEEauthorrefmark{2}\,\orcidlink{0000-0002-3829-3545},~\IEEEmembership{Senior Member,~IEEE},\\
        Aboelmagd Noureldin\IEEEauthorrefmark{1,2}\,\orcidlink{0000-0001-6614-7783},~\IEEEmembership{Senior Member,~IEEE}
    }
            \thanks{\IEEEauthorrefmark{1}Dept. of Electrical and Computer Engineering, Queen's University at Kingston, ON, K7L3N9, Canada, and the Navigation and Instrumentation (NavINST) Lab at the Royal Military College of Canada, Kingston, ON, K7K7B4, Canada. Emails: \texttt{hany.ragab@queensu.ca, aboelmagd.noureldin@rmc-cmr.ca}}
        
        \thanks{\IEEEauthorrefmark{2}School of Computing, Queen's University at Kingston, ON, K7L2N8, Canada. Emails: \texttt{\{sidney.givigi, nourelda\}@queensu.ca}}
}

\maketitle
\newcommand{\nint}[1]{\ensuremath\left\lfloor#1\right\rceil}

\begin{abstract}
Modern on-road navigation systems heavily depend on integrating speed measurements with inertial navigation systems (INS) and global navigation satellite systems (GNSS). Telemetry-based applications typically source speed data from the On-Board Diagnostic II (OBD-II) system. However, the method of deriving speed, as well as the types of sensors used to measure wheel speed, differs across vehicles. These differences result in varying error characteristics that must be accounted for in navigation and autonomy applications. This paper addresses this gap by examining the diverse speed-sensing technologies employed in standard automotive systems and alternative techniques used in advanced systems designed for higher levels of autonomy, such as Advanced Driver Assistance Systems (ADAS), Autonomous Driving (AD), or surveying applications. We propose a method to identify the type of speed sensor in a vehicle and present strategies for accurately modeling its error characteristics. To validate our approach, we collected and analyzed data from three long real road trajectories conducted in urban environments in Toronto and Kingston, Ontario, Canada. The results underscore the critical role of integrating multiple sensor modalities to achieve more accurate speed estimation, thus improving automotive navigation state estimation, particularly in GNSS-denied environments.
\end{abstract}


\section{Introduction}

Accurate speed estimation, particularly forward velocity, is crucial for automotive applications such as fleet management, navigation, Advanced Driver Assistance Systems (ADAS), and Autonomous Driving (AD) \cite{bathla_autonomous_2022}. Vehicle tracking depends on real-time vehicular data for route optimization, while ADAS and AD require precision for functions like lane-keeping and collision avoidance. Mercedes-Benz's DRIVE PILOT \cite{group_mercedes-benz_2022} emphasizes redundancy and safety, using moisture sensors in the wheel wells to monitor road wetness and backup systems to ensure reliable operation in challenging conditions.

\begin{figure}[t]
    \centering
    \includegraphics[width=\linewidth]{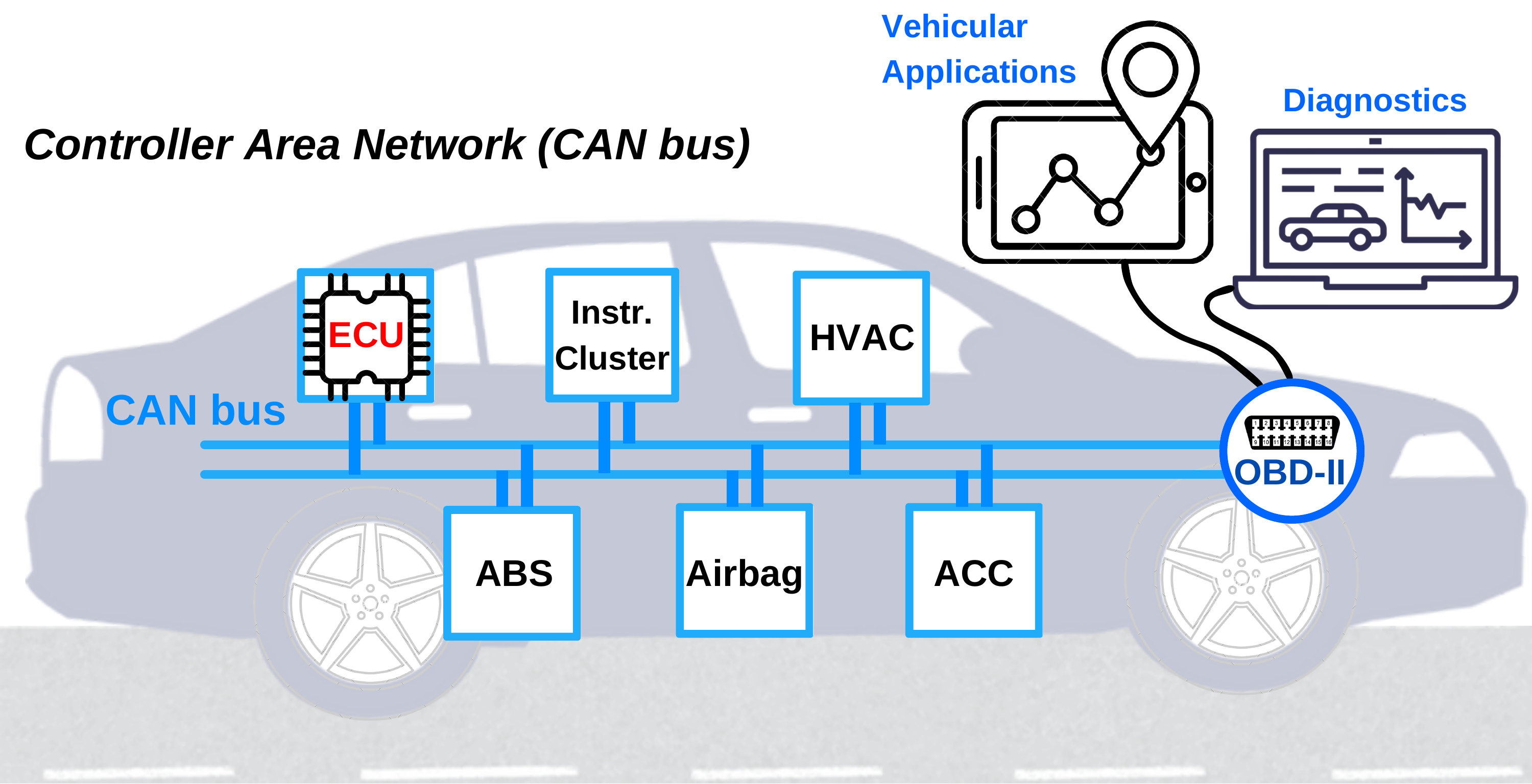}
    \caption{Interconnections of Automotive Sensors and Systems via CAN.}%
    \label{fig:Vehicular_Applications}
\end{figure}

Many complementary systems, such as navigation and driver assistance solutions added post-manufacture, access essential vehicular data through the On-Board Diagnostic II (OBD-II) port \cite{schafer_commute_2018}. As illustrated in Fig. \ref{fig:Vehicular_Applications}, the OBD-II interface connects to the vehicle’s primary network, typically the CAN bus, or other protocols depending on the make and model. It allows communication with subsystems such as the Electronic Control Units (ECUs), Anti-lock Braking System (ABS), Instrument Cluster, and Adaptive Cruise Control (ACC) modules, to name a few. This standardized interface provides critical parameters, including engine RPM and wheel speed, facilitating seamless integration with internal and external systems. However, the sensors and techniques used to derive speed measurements can differ significantly between land vehicles. These differences can introduce variations in error characteristics, which can profoundly impact the reliability and accuracy of downstream systems relying on this data.

Wheel speed, in particular, plays a critical role in enhancing Inertial Navigation Systems (INS) and enabling the Reduced Inertial Sensor System (RISS), a simplified INS variant that predominantly depends on external speed inputs for operation \cite{iqbal_integrated_2008}. The quality of wheel speed data is essential in GNSS-denied environments or, in the case of modern systems, signals of opportunity (SOPs)-denied conditions, where errors in proprioceptive measurements can propagate and degrade dead-reckoning performance. Furthermore, a limited understanding of sensor types, error characteristics, and data formats can compromise the accuracy and reliability of complementary systems. The key contributions of this paper are:

\begin{figure*}[t]
\centering
\includegraphics[width=0.99\textwidth]{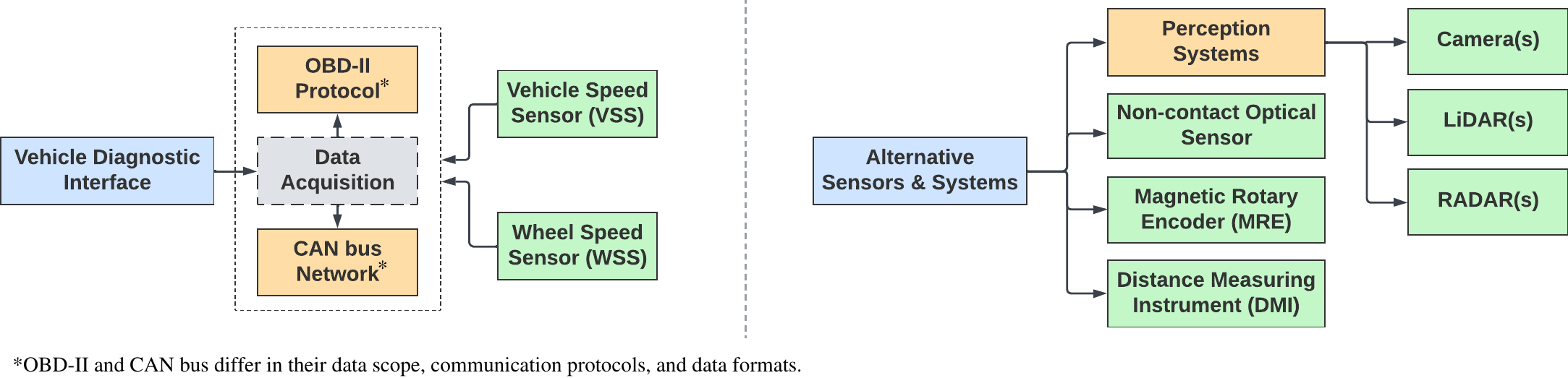}
\vspace{-0.2cm}
\caption{Methods for estimating the speed of a land vehicle without relying on GNSS or IMU integration.}
\label{fig:veh_speeds}
\end{figure*}

\begin{enumerate}[label=\arabic*)]
    \item Comprehensive overview of automotive speed sensing: We analyze methods for estimating vehicle speed, highlighting classical and modern sensor technologies.

    \item Comparison of speed derivation processes: 
    We investigate the speed derivation methods employed at the Original Equipment Manufacturer (OEM) level for different vehicle types. We also emphasize the distinctions between OBD-II and CAN bus protocols' data format. 

    \item Real-road data collection and analysis: 
    We utilize road test data to identify speed sensor types from OBD-II-reported messages, compare the accuracy of OBD-II-derived speed estimates with perception-based methods, and conduct an error variance analysis as a step toward achieving effective sensor fusion.
\end{enumerate}

The remainder of this paper is organized as follows: 
Section \ref{sec:Related_Work} reviews related work on ground vehicle speed sensing and its applications in navigation. 
Then, Section \ref{sec:OBDII_Vehicle_Speed_Derivation} explores OBD-II communication, data formatting, and speed derivation for various vehicle configurations, along with an overview of wheel speed sensors (WSS). 
Following this, Section \ref{sec:exp_work_and_results} presents experimental results, including sensor identification, error variance analysis, and the integration of stereo cameras for complementary speed estimation. Finally, Section \ref{sec:Conclusion} concludes the paper and outlines future research directions.

\section{Related Work} \label{sec:Related_Work}

Several publicly available automotive datasets offer valuable insights for research and development; however, only a few include data collected directly through a vehicle's proprietary diagnostic interface. Among datasets that provide speed data, high-resolution Distance Measuring Instruments (DMIs), such as those featured in the DARPA dataset \cite{huang_high-rate_2010} and the Ford Campus dataset \cite{pandey_ford_2011}, are commonly used. While these systems deliver exceptional accuracy, their high cost and complexity often exceed the requirements of low-cost automotive navigation applications.

In addition to built-in vehicular speed sensing and DMIs, alternative systems have been explored for specialized ground vehicle testing. Magnetic Rotary Encoders (MREs) \cite{jeong_complex_2019} and Non-contact Optical Sensors \cite{koschorrek_multi-sensor_2013,zhang_gnssmulti-sensor_2024} are examples used in extreme conditions like snow, ice, or wet surfaces, or in harsh environments with dust, moisture, extreme temperatures, and mechanical shocks. Despite their robustness and precision, these niche applications and high costs limit their practical use in everyday automotive systems.

Datasets like nuScenes \cite{caesar_nuscenes_2020}, A2D2 \cite{geyer_a2d2_2020}, and \texttt{comma2k19} \cite{schafer_commute_2018} aim to make built-in vehicular data more accessible. These datasets rely on the CAN bus network, which poses implementation challenges due to the lack of standardization across manufacturers. Each manufacturer defines unique CAN message structures and IDs, typically documented in \texttt{DBC} files. Consequently, interpreting data such as vehicle wheel speed from the CAN bus often requires extensive knowledge of the vehicle's architecture or reverse engineering. In contrast, the OBD-II protocol provides a standardized and interoperable approach for extracting diagnostic data, including vehicle speed, across all OBD-II-compliant vehicles. Table \ref{tab:comparison_obd_can} summarizes the key differences between the OBD-II protocol and CAN bus. Although the OBD-II protocol is straightforward, the diverse methods used for deriving vehicle speed from built-in sensors remain underexplored in the literature, creating significant challenges for navigation system integrators aiming to develop universally compatible, robust, and low-cost solutions.
\vspace{-0.3cm}

\begin{table}[ht]
  \begin{threeparttable}[b]
    \caption[Comparison between OBD-II and CAN Bus]{Comparison between OBD-II protocol and CAN Bus}
    \label{tab:comparison_obd_can}
    \centering
    \begin{tabularx}{\linewidth}{|>{\raggedright\arraybackslash}p{0.24\linewidth}|>{\raggedright\arraybackslash}p{0.28\linewidth}|>{\raggedright\arraybackslash}X|}
      \hline
      \textbf{Aspect} & \textbf{OBD-II Standard} & \textbf{CAN Bus}\tnote{\textcolor{blue}{\ddag}} \\
      \hline
      Purpose & Diagnostics and emissions monitoring & Real-time communication between ECUs \\
      \hline
      Standardization & Standardized PIDs\tnote{\textcolor{blue}{\dag}} (e.g., \texttt{0x0D}) & Manufacturer-specific (i.e., CAN ID varies) \\
      \hline
      Data Access & Limited to subset of vehicle parameters & Full access to all CAN messages \\
      \hline
      Communication Model & Request-Response & Broadcast (continuous streaming) \\
      \hline
      Speed of Access & Slower, query-based & Faster, real-time \\
      \hline
      Physical Access & OBD-II port & OBD-II port or direct CAN wiring \\
      \hline
    \end{tabularx}
    \begin{tablenotes}
      \item[\textcolor{blue}{\dag}] PID: Parameter ID as defined in SAE J1979.
      \item[\textcolor{blue}{\ddag}] CAN remains the most adopted in-vehicle network; Ethernet (802.3bw) offers similar features and higher data rates for emerging applications.
    \end{tablenotes}
  \end{threeparttable}
\end{table}

In addition to the often omitted derivation process, the navigation and instrumentation literature offers limited clarity on the specific sensors automotive suppliers use for speed estimation. Researchers often overlook the type of speed sensors employed in their test vehicles, commonly employ the general term \textit{odometer} which covers a wide variety of instruments, and frequently rely on traditional Inertial Measurement Unit (IMU) and GNSS error modeling techniques for sensor fusion \cite{iqbal_integrated_2008, noureldin_fundamentals_2013, chang_gnssimuodolidar-slam_2020,karamat_enhanced_2015}. While these approaches have been widely used, they may not guarantee optimal estimation of vehicle speed. Therefore, identifying and characterizing vehicular speed sensors is critical for designing robust and effective navigation systems.

Fig. \ref{fig:veh_speeds} categorizes vehicle speed-sensing technologies into two primary groups: built-in diagnostic data and alternative sensors and systems. The first group includes OBD-II-derived speed measurements and CAN bus-derived data from Vehicle Speed Sensors (VSS) and WSS, both of which are widely integrated into vehicles’ onboard systems. The second group consists of external sensors, such as non-contact optical sensors, DMIs, and perception-based systems like cameras, LiDARs, and RADARs. Cameras and RADARs, in particular, are increasingly utilized not only for ADAS but also for forward velocity estimation. Recent studies have demonstrated the effectiveness of cameras for ego-motion tracking \cite{ferrari_stereo_2018} and RADARs for ego-velocity estimation \cite{de_araujo_toward_2024}. These perception-based approaches offer practical and cost-effective alternatives, particularly in scenarios where traditional automotive speed sensors are limited or unavailable.

\section{OBD-II Vehicle Speed Derivation} \label{sec:OBDII_Vehicle_Speed_Derivation}

Modeling errors in consumer-level OBD-II-derived speed measurements presents a significant challenge due to the restricted data available through the vehicle's regular OBD-II protocol. Generally, the OBD-II standard provides data on the overall vehicle speed rather than the speed of each individual wheel. Additionally, the speed data is often in \verb|integer| format, which lacks the resolution of decimal points found in \verb|double| or \verb|float| types commonly available in CAN bus systems, leading to less precise measurements. When querying the OBD-II system for vehicle speed, a command is sent, and the system responds with encoded data. Below is an example illustrating this process: \vspace{2mm}
{
\setlength{\leftmargini}{1.5em} 
\begin{quote}
\begin{lstlisting}[caption={OBD-II Command and Response Example.}, label={lst:obd_example}]
# Command Sent
> 010D
# Response Received
41 0D 3C
(*@\textit{Interpretation:}@*)
# 41: Response for mode 01
# 0D: Echo of the PID for vehicle speed
# 3C: Vehicle speed in hexadecimal (60 in decimal, km/h)
\end{lstlisting}
\end{quote}
}

The command \texttt{010D} represents a request for the current vehicle speed (PID \texttt{0x0D}). The response \texttt{41 0D 3C} indicates the vehicle speed as \texttt{0x3C}, which is \texttt{60} $km/h$ when converted to decimal\footnote{Automakers may configure ECUs to report diagnostic data in region-specific units. For example, vehicle speed may be in miles per hour ($mph$) for the USA or in kilometers per hour ($km/h$) for Canada and Europe.}. In other words, the reported speed through standard OBD-II interfacing is:

\begin{equation}
V_{\text{v}}^{\text{OBD}} = \nint{\hat V}.
\end{equation}

Here, \(\nint{\cdot}\) denotes the operator for truncating to the nearest integer, and \(\hat{V}\) represents the estimated speed of the vehicle, regardless of the computation method. There are two primary approaches to compute vehicle speed from built-in sensors: one utilizes transmission data, and the other leverages WSS data, readily available in ABS-equipped vehicles.

\subsection{Transmission-based Vehicle Speed Determination} \label{sec:Transmission-based_Vehicle_Speed_Estimation}

\begin{figure}[htb]
    \centering
    \includegraphics[page=1,width=0.9\linewidth]{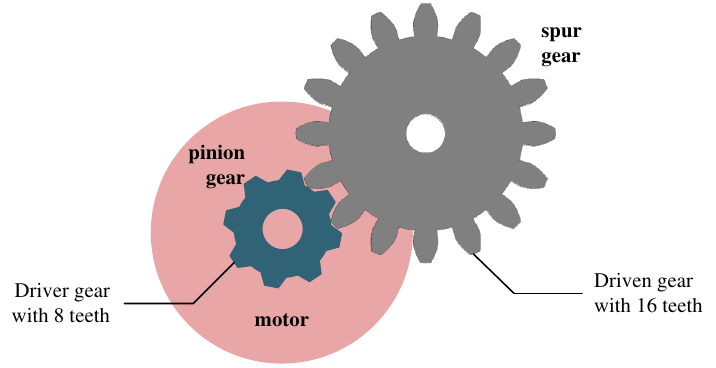}
    \caption{Illustration of gear ratio.}
    \label{fig:gear_ratio}
\end{figure}

The estimation of vehicle speed (\(V_\text{v}\)) from transmission data relies on the relationship between engine speed in RPM and drivetrain parameters, especially gear ratio. The gear ratio governs the conversion of engine speed (input) to wheel speed (output). For example, in Fig. \ref{fig:gear_ratio}, a driving gear with 8 teeth and a driven gear with 16 teeth creates a 2:1 ratio, where the output rotates at half the input speed. This adjustment is crucial for controlling speed and torque. The gearbox alters engine speed based on the engaged gear ratio (\(i_x\)), which changes with each gear shift, while the differential uses its gear ratio (\(i_0\)) to balance wheel speeds, critical for smooth turning.

\vspace{-0.3cm}

\begin{figure}[htb]
    \centering
    \includegraphics[width=0.6\linewidth]{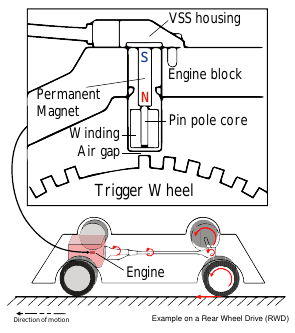}
    \vspace{-0.4cm}
    \caption{Vehicle Speed Sensor (VSS).}%
    \label{fig:asset3}
\end{figure}

In non-ABS vehicles, where speed is derived from the engine's VSS (see Fig. \ref{fig:asset3}), the calculation for speed in \( m/s \) is given by the following formula:

\begin{equation} 
V_{\text{v}}^{\text{OBD}} = \nint{\left({\frac{N_e \cdot \pi \cdot r_w}{30 \cdot i_x \cdot i_0}}\right)},
\end{equation}

\(N_e\) represents the engine speed in RPM, typically measured by a crankshaft position (CKP) sensor or an engine speed sensor. The wheel radius (\(r_w\)), measured in meters (\(m\)), converts the rotational speed of the wheels into linear speed. This calculation assumes a slip-free mechanical connection between the engine and wheels, ensuring efficient power transfer and accurate speed estimation.
\vspace{-0.3cm}

\subsection{ABS-based Vehicle Speed Determination} \label{sec:ABS-based_Vehicle_Speed_Estimation}

\begin{figure}[htb]
    \centering
    \includegraphics[width=0.6\linewidth]{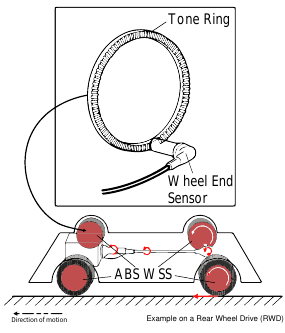}
    \vspace{-0.4cm}
    \caption{Wheel Speed Sensor (WSS).}%
    \label{fig:asset4}
\end{figure}

The ABS WSS, depicted in Fig. \ref{fig:asset4}, produces output signals as the wheel turns, i.e. \textit{wheel ticks}. These signals are then transmitted to the vehicle's ECU or a specific ABS control module. The main role of the ECU or ABS module is to interpret these signals to determine each wheel's speed.

The compute module calculates the speed of each individual wheel using the formula:

\begin{equation}
V_{\text{w}} = \frac{2 \pi \cdot r_w \cdot N_p}{N_t \cdot \Delta t},
\end{equation}

where $V_{\text{w}}$ denotes the linear speed of an individual wheel ($m/s$), $N_p$ is the number of pulses detected during the sampling interval $\Delta t$ ($s$), and $N_t$ is the total number of teeth on the reluctor ring.
To determine the vehicle's overall speed, the ECU averages the rotational speeds of all the individual wheels. This average is then used for various control and monitoring functions within the vehicle's ABS and electronic stability control (ESC) systems.

Compared to the transmission-based system, ABS-based speed estimation offers greater robustness and accuracy while involving fewer parameters. 
Thus, mathematically, the actual ABS wheel speed reported through the regular OBD-II protocol is computed as follows:

\begin{equation}
V_{\text{v}}^{\text{OBD}} = \nint{\left(\frac{V_{\text{w}_1} + V_{\text{w}_2} + V_{\text{w}_3} + V_{\text{w}_4}}{4}\right)},
\end{equation}

where $V_{\text{w}_i}$ denotes the velocity of the $i^{th}$ wheel ($m/s$), assuming each wheel is equipped with a WSS. To avoid reliance on automotive OEM collaboration or specialized tools (e.g., CAN bus reverse engineering) for accessing detailed sensor information in test vehicles, navigation and mapping entities often rely on carefully installed and calibrated DMIs.

\subsection{Wheel Speed Sensor Types}

When dealing with WSS, limited information can be found in the navigation and instrumentation literature related to the type of WSS automotive suppliers use and their corresponding accuracy. Within the automotive industry, two primary types of WSS are currently utilized, each possessing unique advantages and disadvantages. These are referred to as $passive$ and $active$ wheel speed sensors, respectively. The operational mechanisms of these sensors exceed the scope of this paper, however, their main differences are shown in Table \ref{tab:wss_automotive_comparison}. 
\vspace{-0.1cm}
\begin{table}[htb]
\caption{Automotive WSS: A comparison.}
\label{tab:wss_automotive_comparison}
  \centering
  \begin{tabular}{lll}
    \toprule
    \multicolumn{3}{c}{WSS Types} \\
    \hline
     & \textit{Passive} & \textit{Active} \\
    \midrule
    Advantages &
    - Lower cost & - Better accuracy at  \\  & & \hspace{2mm}low speeds \\ 
    & - Easy to maintain & - Suitable for high-\\ & & \hspace{2mm}performance applications \\
    & - No power supply required & - Advanced diagnostic \\ & & \hspace{2mm}capabilities \\ [.5\normalbaselineskip] 
    \midrule
    Drawbacks &
    - Poor accuracy at & - Higher cost \\ & \hspace{2mm}low speeds  & - More complex design \\
    & - Limited diagnostic  &  - More susceptible to \\ & \hspace{2mm}capabilities & \hspace{1mm} electromagnetic interference \\ \bottomrule

  \end{tabular}
\end{table}

While Passive WSS (PWSS) are simple in terms of design and operation, passive sensors may be less accurate, especially at lower speeds. In fact, the ECU typically disregards speed signals below a certain speed threshold ($\leq$ 3 $km/h$) to prevent false triggering from signal noise.  In contrast, active WSS can detect speeds as low as $0.1$ $km/h$ \cite{reif_brakes_2014} and deliver consistent error characteristics across the vehicle speed spectrum under ideal conditions due to their advanced design and operation.

\section{Experimental Work and Results} \label{sec:exp_work_and_results}
This section describes the experimental setup, details of the road tests, and results from three real-road trajectories performed in Kingston and Toronto, Ontario, Canada.

\subsection{Experimental Setup}

\begin{figure}[htb]
    \centering
    \includegraphics[width=1\linewidth]{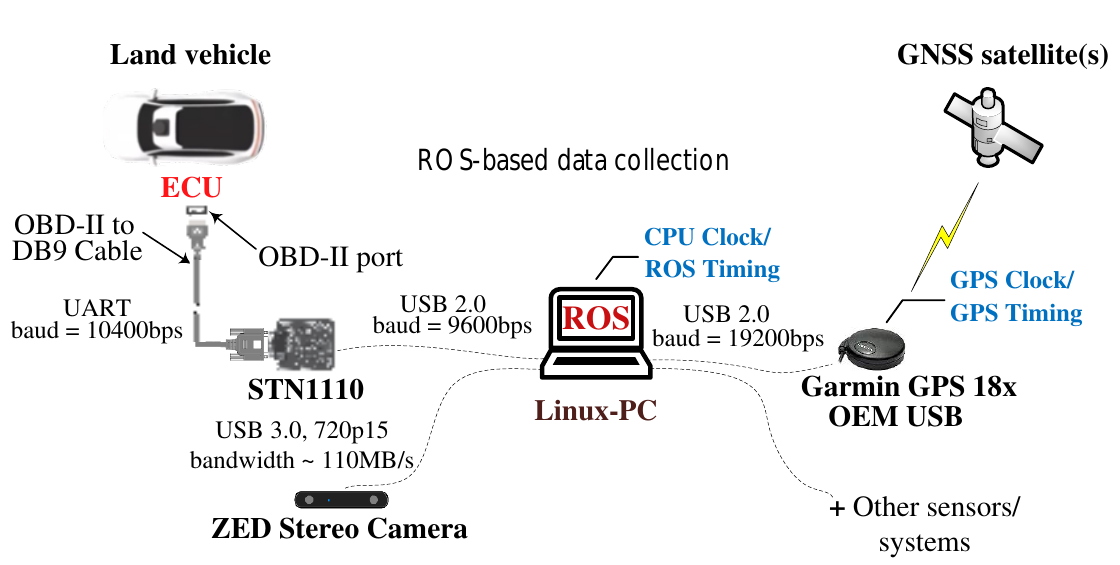}
    \vspace{-6mm}
    \caption{Data logging setup.}%
    \label{fig:obdii_logging_speed}
\end{figure}

The experiments were conducted using an ABS-equipped Toyota Sienna minivan, with the data logging setup outlined in Fig. \ref{fig:obdii_logging_speed}. A detailed dataset description is provided in \cite{ragab_integration_2024}, and supporting materials are publicly available\footnote{GitHub repository: \url{https://github.com/hanymragab/plans2025-resources}}.
\vspace{-4mm}

In this setup, we employed the \texttt{STN1110} OBD-to-UART interpreter to interface with the vehicle's OBD-II, enabling the logging of speed data using the Robot Operating System (ROS) on a Linux PC. Additionally, camera frames were captured using a ZED StereoLabs camera. To ensure precise synchronization, a low-cost Garmin 18x OEM USB module was used to align the Linux PC's clock with a high-end tactical grade NovAtel ProPak6$^{\text{TM}}$ GNSS/IMU unit, which served as the source of ground truth data. The following section details the key findings and insights from the experiments.

\subsection{Results and Discussion}

\subsubsection{Identification and Error Modeling of Automotive Speed Sensors}
In integrated navigation systems, understanding how the integrity of sensed measurements impacts overall navigation performance is essential. Consequently, performing error variance analysis on logged vehicle speed measurements is a critical step in evaluating sensor accuracy. To achieve reliable error variance estimates, real-world data must be collected from multiple road test trajectories conducted with the same vehicle. The data from each trajectory can then be post-processed to estimate the average error variance as a function of a quantity known to influence measurement accuracy.

This average error variance can be modeled using a suitable mathematical function, such as a polynomial or exponential fit, to characterize its behavior with respect to the influencing quantity. Such modeling provides valuable insights into the error characteristics, enabling more informed decisions for the design and optimization of sensor fusion algorithms and related systems.

\begin{figure}[htb]
    \centering
    \includegraphics[width=0.77\linewidth]{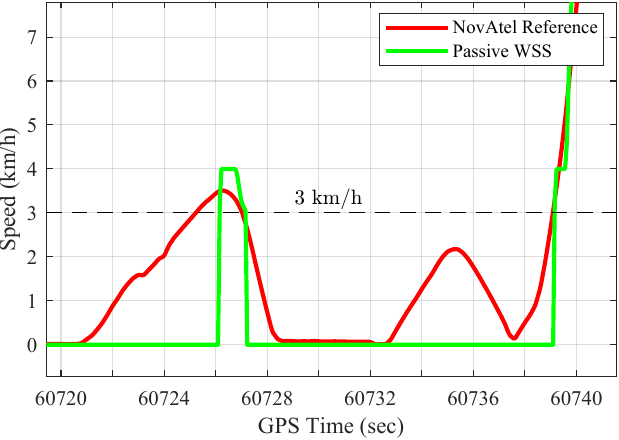}
    \vspace{-3mm}
    \caption{OBD-II-reported vehicle speed (green) as compared to the reference high-end solution (red).}
    \label{fig:pwss_3kmph_th}
\end{figure}

\begin{figure}[htb]
    \centering
    \includegraphics[width=0.93\linewidth]{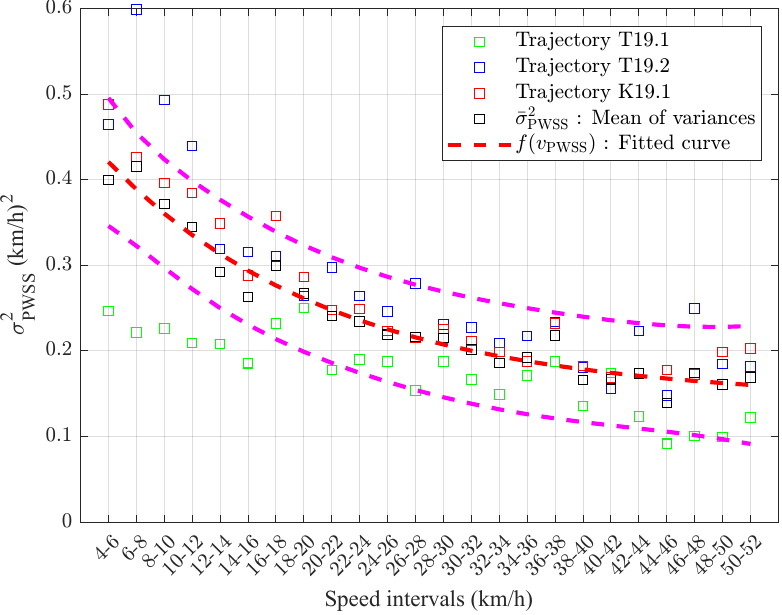}
    \vspace{-3mm}
    \caption{Error variance analysis of passive wheel speed sensor data from three trajectories: ‘T’ and ‘K’ indicate Toronto and Kingston, respectively, while ‘19’ refers to data collected in 2019.}%
    \label{fig:pwss_variances}
\end{figure}

To identify the type of WSS, we first examined the speed profiles for values below 3 $km/h$. If such readings were present, the sensor was classified as \textit{active}; otherwise, it was considered \textit{passive}. This approach aligns with the known behavior of passive WSS, which typically do not report speeds below this threshold. A sample speed profile from one of the test trajectories (Fig.~\ref{fig:pwss_3kmph_th}) shows no data in the low-speed range, indicating the use of passive sensors.
Given that passive WSS accuracy improves with increasing speed, the error variance analysis was conducted as a function of travel speed. Fig.~\ref{fig:pwss_variances} presents results from Kingston and Toronto, which reflect this trend and were used to fit the variance model.

The vehicular speed measurements were directly compared against the resultant velocity computed from the high-end tactical-grade integrated navigation solution. After further data analysis, a 2$^{nd}$ degree exponential function was selected in the data fitting process as it provides a flexible and efficient way to model the data accounting for non-linearities. The 2$^{nd}$ degree exponential function takes the form of:

\begin{equation}
  \sigma^2_{PWSS}(v_{f}) = ae^{bv_{f}} + ce^{dv_{f}},
  \label{eqn:sigma_squared_PWSS}
\end{equation}

where the coefficients $a$, $b$, $c$, and $d$, are determined using an optimization algorithm. In this case, the obtained function is $\sigma^2_{PWSS}(v_{f}) = 0.3095e^{-0.1241v_{f}} + 0.1477e^{-0.0009149v_{f}}$ and the coefficients are determined with 95\% confidence bounds. This means that there is a 95\% probability that the actual values of the coefficients fall within the bounds of the confidence intervals which corresponds to $\pm 2\sigma$. The obtained function provides a good representation of the data and can be used for further analysis or predictions.

\begin{figure}[htb]
    \centering
    \includegraphics[width=0.84\linewidth]{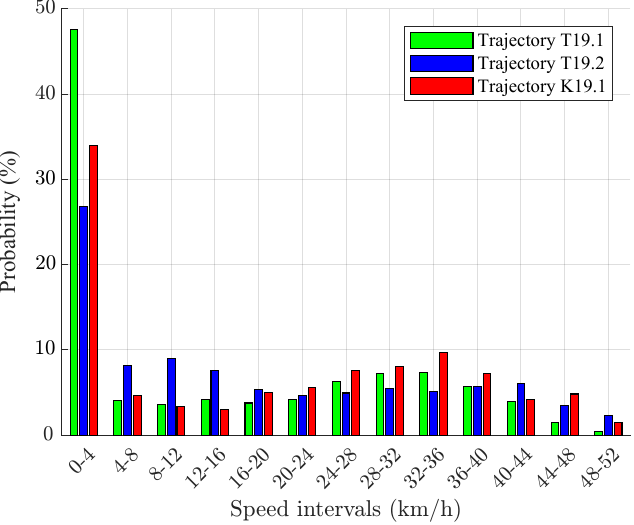}
    \vspace{-2.2mm}
    \caption{Histogram-based Probability Mass Function (PMF) in terms of speed intervals for three real road trajectories, two of which are in Toronto.}%
    \label{fig:pdf_speeds}
\end{figure}

Driving patterns vary by environment, with vehicles typically moving slower in dense urban areas and downtown cores than on highways. As shown in Fig.~\ref{fig:pdf_speeds}, the histogram-based probability mass function (PMF) shows that speeds between 0–4~km/h account for 35–45\% of samples in urban trajectories T19.1 and K19.1. In contrast, for T19.2, which includes highway segments, this falls below 30\%. These trends highlight the need to assess perception-based speed estimation in GNSS-challenging urban settings, where passive WSS may fail to capture low-speed motion. 

\subsubsection{OBD-II-derived speed vs Stereo VO-derived speed} 
\begin{figure}[htb]
    \centering
    \includegraphics[width=0.85\linewidth,]{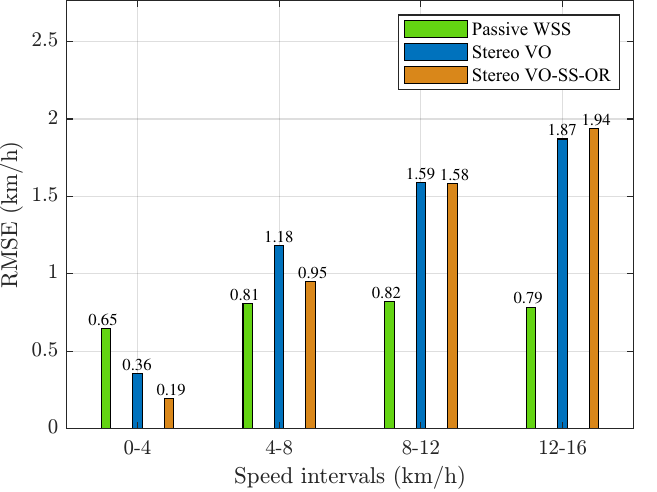}
    \vspace{-3mm}
    \caption{Speed RMSE comparison between PWSS and Stereo VO with and without SS-OR.}%
    \label{fig:speed_rmse_sensors}
\end{figure}

To evaluate the performance of speed estimates across different intervals, we compare OBD-II-derived vehicle speeds with results from a stereo visual odometry (Stereo VO) routine and an enhanced Stereo VO algorithm incorporating semantic segmentation-based outlier rejection (Stereo VO-SS-OR), as introduced in our earlier work \cite{ragab_utilization_2020}. 
Fig. \ref{fig:speed_rmse_sensors} reveals significant variations in the accuracy of speed measurements. In the lower speed range (0–4 $km/h$), both Stereo VO algorithms outperform the OBD-derived PWSS speed measurements, with Stereo VO-SS-OR achieving the lowest root mean square error (RMSE), unlike at higher speed intervals. 

\begin{figure}[htb]
    \centering
    \includegraphics[width=0.95\linewidth]{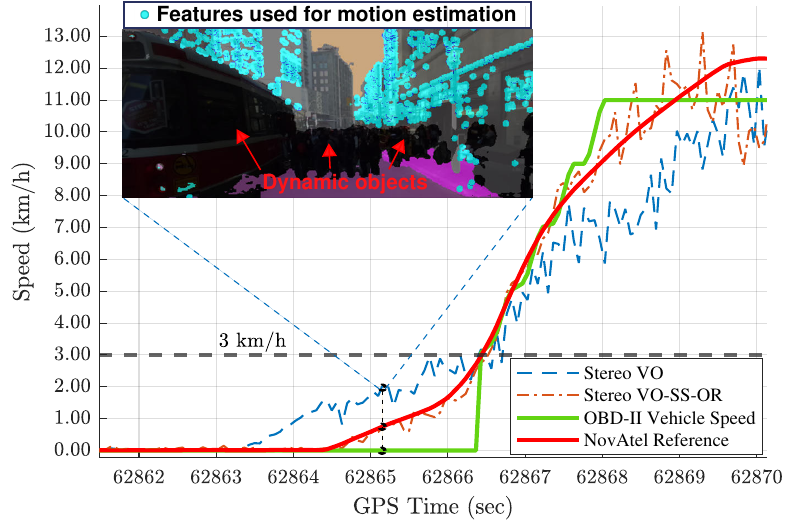}
    \vspace{-3mm}
    \caption{Performance at low travel speeds.}%
    \label{fig:SS_OR_perf}
\end{figure}

Fig. \ref{fig:SS_OR_perf} presents a comparison of speed estimates from Stereo VO, with and without the SS-OR enhancement, alongside OBD-II-derived speeds and the NovAtel reference. It highlights the variability in Stereo VO performance across different scenarios and showcases the capability of the SS-OR-enhanced version to improve vehicular speed estimation, particularly when PWSS fails to provide data. The integration of SS-OR significantly enhances ego-motion estimation accuracy, resulting in more consistent and precise speed measurements, especially at lower speeds of travel in highly dynamic environments.

\begin{figure}[htb]
    \centering
    \includegraphics[width=0.92\linewidth]{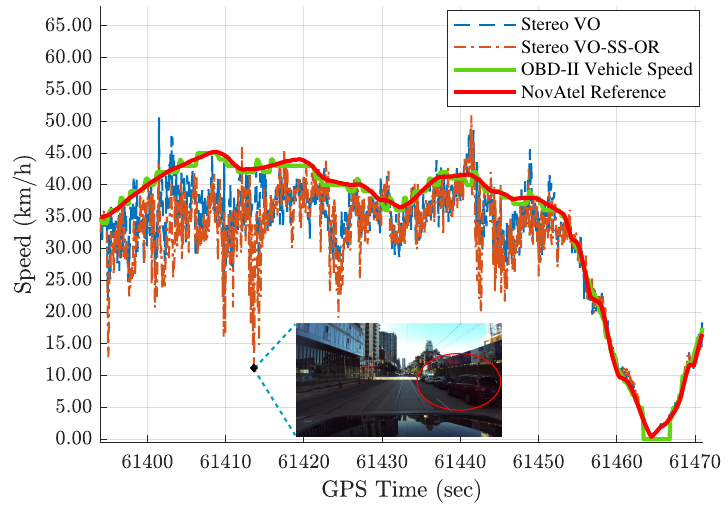}
    \vspace{-2mm}
    \caption{Degeneration of SS-OR: Parked or stationary vehicles misclassified as dynamic, with suboptimal travel speed (vs framerate) and illumination.}%
    \label{fig:SS_OR_degen}
\end{figure}

On the other hand, Fig. \ref{fig:SS_OR_degen} illustrates the challenges faced in low-traffic environments dominated by static objects, which significantly affect the approach’s effectiveness. 
In such scenarios, Stereo VO generates noisier and less accurate speed estimates compared to OBD-II-derived speeds. The performance of Stereo VO-SS-OR deteriorates further at higher travel speeds, primarily due to baseline discrepancies between frames combined with framerate limitations.

Additionally, the abundance of parked cars (as highlighted in the figure), often being misclassified as dynamic within the SS-OR routine, disrupts temporal feature tracking, and impairs accurate ego-motion estimation.
Given these differences, implementing a fusion scheme is essential to selectively weight measurements based on WSS speed data, enabling a more accurate estimation of forward speed. Additionally, this highlights the critical need for integrating object tracking or supplementary sensors, such as RADAR sensors to improve stationary and dynamic object classification, enhancing the overall performance of the SS-OR framework.

\section{Conclusion} \label{sec:Conclusion}
In conclusion, this paper sheds light on a critical yet underexplored aspect of navigation and instrumentation: the diverse sensors and methods for deriving automotive speed measurements from the OEM perspective. By identifying and analyzing these approaches, we demonstrated the strengths and limitations of using OBD-II-derived speed data for land vehicle navigation. Through experiments conducted on three long real road trajectories, we identified the WSS type employed and showcased the advantages of using alternative ADAS sensors, such as front-facing stereo cameras, for speed estimation.

While each sensor modality offers unique advantages and limitations, our findings underscore the importance of a fusion engine capable of reliably integrating diverse data sources to ensure robust and resilient navigation performance. Achieving universal plug-and-play capability would require collecting OBD-II-derived speed data across a broad spectrum of vehicles, including those equipped with active wheel speed sensors. This expanded dataset would support the development of more accurate and vehicle-specific error profiles, enabling a more adaptive and generalizable fusion framework that leverages low-cost exteroceptive sensors.

Looking ahead, for more advanced and vehicle-specific systems, reverse engineering in-vehicle communication protocols to access individual wheel speed data directly presents a promising direction. This approach can significantly improve the accuracy of speed estimation and, in turn, enhance the overall performance of the integrated navigation system.

\bibliographystyle{IEEEtran}
\bibliography{final_references}

\end{document}